# An Architecture of Active Learning SVMs with Relevance Feedback for Classifying E-mail

Md. Saiful Islam and Md. Iftekharul Amin

| | |
|---|---|
| Dept. of Computer Science & Engineering | Dept. of Computer Science & Engineering |
| State University of Bangladesh | IBAIS University |
| 77, Satmasjid Road, Dhanmondi | House#05, Road #16, Dhanmondi R/A |
| Dhaka-1205, Bangladesh | Dhaka-1205, Bangladesh |

E-mail: sohel_csdu@yahoo.com and shatil_csdu@yahoo.com

**Abstract**

In this paper, we have proposed an architecture of active learning SVMs with relevance feedback (RF) for classifying e-mail. This architecture combines both active learning strategies where instead of using a randomly selected training set, the learner has access to a pool of unlabeled instances and can request the labels of some number of them and relevance feedback where if any mail misclassified then the next set of support vectors will be different from the present set otherwise the next set will not change. Our proposed architecture will ensure that a legitimate e-mail will not be dropped in the event of overflowing mailbox. The proposed architecture also exhibits dynamic updating characteristics making life as difficult for the spammer as possible.

**Major Area:** Machine Learning, Artificial Intelligence, and Network Security

## 1. Introduction

Due to the rapid development of WWW and Internet, e-mail becomes a very efficient, convenient and shortcut way of communication. But at the same time despite e-mail's numerous advantages, it has become the victim of abuse. The mass posting of unsolicited bulk commercial e-mail has become an increasingly large problem. This unsolicited and unwanted e-mail is known as Spam in the literature. This increase in Spam is annoying. An interesting aspect of this rise in spam is that unique spam attacks are also on the rise. More about spam will be found in http://www.junke-mail.org, http://spam.abouse.net and http://cauce.org. There have been various learning machines to classify e-mail [7][8][11].

Solutions to the proliferation of spam includes either technical or regulatory [1]. Technical solutions include filtering based on sender address or header content. Microsoft junk and adult content e-mail filters work by looking for key words emphasizing on which words the filter will look for and where [2]. Although sender address based filtering may be useful but content based filtering may sometimes block valid messages [3][4]. Since missing legitimate e-mail is an order of magnitude worse than receiving spam, we have no intention to automatically reject e-mail that is classified as spam. Rather, we will allow user to mark (label) their e-mail as either spam or nonspam. After a finite number of examples are collected, the learning machine will trained and performance on new examples will be predicted. Our proposed mail server will deliver e-mail to the user in decreasing order of probability that the e-mail is nonspam. It is then up to the user to either read the e-mail or trash the e-mail. He or she can also provide feedback to the mail server by indicating if an e-mail misclassified. This is what we want to define relevance feedback (RF). Then we will modify our classifier to take into account this. In the case of overflowing mailbox to provide rooms for less probable spam mail that has the high probability to be legitimate we will delete mails at the bottom of the list. This is one of the promising characteristics of our architecture.

Our paper is organized as follows: Section 2 will describe recent trends and techniques used to classify e-mail using SVMs, Section 3 will describe briefly about SVMs, Section 4 will describe active learning strategy, Section 5 will describe several design choices and issues and References are listed in Section 7.

## 2. Related Works

The success of support vector machine (SVM) in solving real-life problems made it not only a tool for the theoretical analysis but also a tool for developing algorithms for real-world problems [5][7][8][9][12]. Their efficiency in providing solutions to classification and function approximation problems is the ongoing research issue. Our intention is to classify e-mail using support vector machines. Since this is a new type of research in the area of machine learning, much of the works are still undone. Drucker *et. al.* have described SVMs for classifying e-mail in [8] and showed that SVMs perform best among different learning algorithms. They emphasized on rank ordering of the e-mail rather than rejecting it. But since their learner learns only once, it will not be able to cope with newer strategies adopted by spammer that can fake their classifier [4]. To take account into this problem our classifier will update itself upon receiving relevance feedback from users if any e-mail misclassified. Therefore it will be almost impossible for the spammer to misguide our classifier. The concept of relevance feedback we used here is as same as relevancy feedback (RF) in information retrieval (IR) of text documents [9]. The advantages of support vector machines with active learning strategy for spam classification has been described in [7]. But though their architecture exhibits active learning strategy, it is incomplete in the above sense that we claimed for [8]. In this paper we have described a complete architecture of active learning SVMs with relevance feedback for classifying e-mail that is free from of the above problems. It also presents other promising advantages that may prove it as a unique candidate.

## 3. Support Vector Machines

The key concepts of SVMs are the following: there are two classes, $y_i \in \{-1, 1\}$ and there are $N$ labeled training examples:

$$(x_1, y_1), (x_2, y_2), ..., (x_N, y_N), x \in R^d$$

where $d$ is the dimensionality of the vector.

If the two classes are linearly separable, then one can find an optimal weight vector $w$ such that $\|w\|^2$ is minimum; and

$$w \bullet x_i - b \geq 1 \quad if \quad y_i = 1;$$
$$w \bullet x_i - b \leq -1 \quad if \quad y_i = -1;$$

or equivalently

$$y_i(w \bullet x_i - b) \leq 1.$$

Training examples that satisfy the equality are termed support vectors. The support vectors define two hyperplanes, one that goes through the support vectors of one class and one goes through the support vectors of the other class. The distance between the two hyperplanes defines a margin and this margin is maximized when the norm of the weight vector $\|w\|$ is minimum. Vapnik has shown we may perform this minimization by maximizing the following function with respect to the variables $\alpha_i$:

$$W(\alpha) = \sum_{i=1}^{N} \alpha_i - \sum_{i=1}^{N}\sum_{j=1}^{N} \alpha_i \alpha_j (x_i \bullet x_j) y_i y_j$$

subject to the constraint: $0 \leq \alpha_j$ where it is assumed there are $N$ training examples, $x_i$ is one of the training vectors, and $\bullet$ represents the dot product. If $\alpha_j > 0$ then $x_j$ is termed a support vector. For an unknown vector $x_j$ classification then corresponds to finding

$$F(x_j) = sign\{w \bullet x_j - b\}$$

where

$$w = \sum_{i=1}^{r} \alpha_i y_i x_i$$

and the sum is over the $r$ nonzero support vectors (whose $\alpha$'s are nonzero).

The advantage of the linear representation is that w can be calculated after training and classification amounts to computing the dot product of this optimum weight vector with the input vector.

For the nonseparable case, training errors are allowed and we now must minimize

$$\|w\|^2 + C\sum_{i=1}^{N} \xi_i$$

subject to the constraint

$$y_i(w \bullet x_i - b) \leq 1 - \xi \quad \xi \geq 0$$

$\xi$ is a slack variable and allows training examples to exist in the region between the two hyperplanes that go through the support points of the two classes. We can equivalently minimize $W(\alpha)$ but the constraint is now $0 \leq \alpha_i \leq C$ instead of $0 \leq \alpha_j$. Maximizing $W(\alpha)$ is quadratic in $\alpha$ and may be solved using quadratic programming techniques.

The advantage of linear SVM is that execution speed is very fast and there are no parameters to tune except the constant C. Drucker *et. al.* [8] has shown that the performance of the SVM is independent of the choice of C as long as C is large enough (over 50). Another advantage of SVM's is that they are remarkably intolerant of the relative sizes of the number of training examples of the two classes. In most learning algorithms, if there are many more examples of one class than another, the algorithm will tend to correctly classify the class with the larger number of examples, thereby driving down the error rate. Since SVM's are not directly trying to minimize the error rate, but trying to separate the patterns in high dimensional space, the result is that SVM's are relatively insensitive to the relative numbers of each class. For instance, new examples that are far behind the hyperplanes do not change the support vectors. The possible disadvantages of SVM's are that the training time can be very large if there are large numbers of training examples and execution can be slow for nonlinear SVM's, neither of these cases being present here [8].

## 4. Active Learning

Pool based active learning involves selecting a training set of examples $T$ from a pool of unlabeled examples $U$. The examples in $T$ are labeled and used for learning. An iterative process involving active learning at each step is called the active learning cycle. This cycle proceeds as follows.

At first, all of the examples in the pool are unlabeled and therefore members of $U$, the pool of unlabeled examples. An initial training set $T_o$ must have at least one active and one inactive example in order for the maximum margin hyperplane to be found. Therefore we generate $T_o$ by randomly selecting examples from $U$ and labeling them until at least one example with each label is in $T_o$. We then remove the new examples from the unlabeled pool by making $U_o = U - T_o$. In turn, we can use active learning to find a new training set $T_1$. $T_1$ contains all the examples from $T_0$, using some selection criterion, we can choose a batch of $N$ examples from $U_o$, label them and place them in set $T_1$. And we get $U_1 = U - T_1$. Repeat this process we can find training set $T_2$ and so on. The iteration of active learning process is:

> Step1. To find $T_o$ by randomly selecting examples in $U$, until one of each
> label is found. $U_o = U - T_o$.
> Step 2. In iteration $i$, select a batch $B_i$ of $N$ examples from $U_{i-1}$ using some
> selection criterion.
> Step 3. Find the label of each example in $B_i$.
> Step 4. $T_i = T_{i-1} \cup B_i$, $U_i = U - T_i$.
> Step 5. Return to step 2 and repeat from step 2 to step 4.

Here we are first given a set of both labeled and unlabeled data. The pool of training examples can be colleted from individual inboxes, thus it will almost impossible for spammer to write spam messages that can beat our e-mail filter [4].

## 5. Design Choices

### 5.1 Feature Representation

A feature is a word. In the development below, $w$ refers to a word, **x** is a feature vector that is composed of the various words from a dictionary formed by analyzing the documents. We take a word as a feature only if it occurs in three or more documents. There is one feature vector per message. **w** refers to a weight vector usually obtained from some combination of the **x**'s. There are various alternatives and enhancements in constructing the vectors [8]. We will use the TF–IDF. TF-IDF uses the TF (term frequency) multiplied by the IDF (inverse document frequency). The document frequency (DF) is the number of times that word occurs in all the documents (excluding words that occur in less than three documents). The inverse document frequency (IDF) is defined as

$$IDF(w_i) = \log\left(\frac{|D|}{DF(w_i)}\right)$$

where $|D|$ is the number of documents. Typically, the feature vector that consists of the TF-IDF entries is normalized to unit length [8].

This kind of representation scheme leads to a very high dimensional feature spaces containing 10000 dimensions or even more [8]. SVMs use overfitting protection mechanism, which does not depend on the number of features. They have potential to handle this kind of problems [7]. Use of stop list consisting of most frequent words like "of", "and", "the", etc. and punctuation marks ".", ",", ";" etc. can be used to lessen the dimensionality [8]. But it is observed that there are more punctuation marks in the legitimate e-mail than in the spam [4]. This indicates that spammers are less interested in good punctuation than the other people. Looking at the most frequent words, it is not obvious which e-mail is spam and which is legitimate. Therefore we concentrated on using all the feature rather than a subset.

### 5.2 Performance Criteria

•*Error Rate*: Error rate is the typical performance measure for two-class classification schemes. However, two learning algorithms can have the same error rate, but the one which groups the errors near the decision border is the better one.

- *False Alarm and Miss Rate*: We define the false alarm and miss rates as

$$miss\ rate = \frac{nonspam...samples...missclassified}{total...nonspam...examples}$$

$$false\ alarm\ rate = \frac{spam...samples...missclassified}{total...spam...examples}$$

The advantage of the false alarm and miss rates is that that they are a good indicator of whether the errors are close to the decision border or not. Given two classifiers with the same error rate, the one with lower false alarm and miss rates is the better one.

### 5.3 Training Examples

The pool of training examples consists of the individual inboxes. Initially there will be no classifier running. Users will receive e-mails whether it is spam or nonspam. But after a certain amount of time users will be requested to send both legitimate e-mail and spam they considered. This colleting process will be continued until pool is sufficient for the learner. When the learning procedure will complete from training examples, our classifier will be activated and users will receive e-mails according to the probability of their e-mail of being spam.

**Selection Methods:** Here we are given a set of both labeled and unlabeled data. The learning task is to assign labels to the unlabeled data as accurately as possible. To construct SVMs, we must construct an optimal classification hyperplane, i.e. the maximal margin hyperplane.
Assume the training data:

$$(x_1, y_1), (x_2, y_2), ..., (x_N, y_N), x \in R^d, y_i \in \{-1, 1\}$$

one subset I for which $y = 1$, and another II subset for which $y = -1$ can be separated by a hyperplane.

If the pattern datasets can be separated with no error, the Euclidian distance between the support vectors and the hyperplane must be the largest, i.e. the margin. The maximal margin classifier for our task can be build based on the following theorem.

***Theorem:*** Given a linearly separable training examples
$$U = \{(x_1, y_1), ..., (x_N, y_N)\}$$
the hyperplane that solve the problem
$$\min\ (w \bullet w)$$
$$\text{subject to } y_i((w \bullet x_i) - b) \geq 1, i = 1, ... N$$

realizes the maximal margin hyperplane with geometric margin $\gamma = 1/\|w\|^2$.

There are few criteria that can be used in each iteration of the active learning cycle as mentioned above. One method is to choose the closest examples to the maximum hyperplane. In step $j$ we select the closest examples to the maximal margin hyperplane of $T_j$. But it will be difficult to perform active learning if there are no unlabeled examples closer to the hyperplane than the support vectors. Another selection method is the furthest selection algorithm, which chooses the furthest examples from the hyperplane. This method is not theoretically motivated, and we expected to perform it poorly.

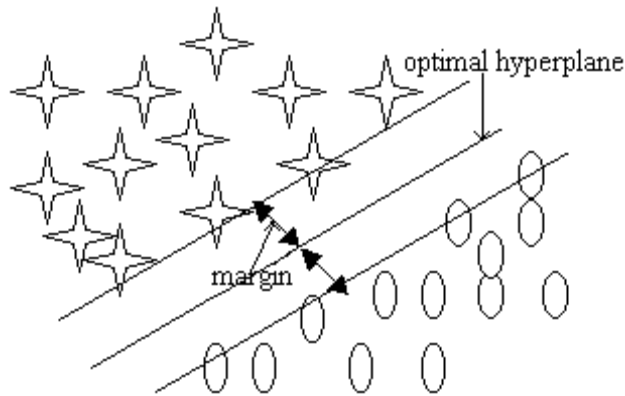

**Fig 1. A maximum margin hyperplane with its support vectors**

### 5.4 Overview of Architecture

The block diagram of our proposed architecture is shown below. The architecture includes a user interface that is responsible for communicating with the user, a pool of training examples, active learning module, e-mail classifier and user mailbox. The total module can be in either Training mode (TM) or Active Mode (AM). In training mode the module will collect individual user e-mails that they considered as spam or legitimate. After collecting sufficient amount of both spam and legitimate e-mails the learning process will begin. Now the module will switch to active mode. In this mode the module classify user's e-mails as spam or nonspam and place them in his or her mailbox according to their probability. When a user observes that an e-mail is misclassified, he or she will provide feedback with misclassified e-mail and its label he or she considered. After receiving feedback the module will switch to training mode and above described procedure will begin again.

### 6. Summary and Further Works

The paper described an architecture of active learning SVMs with relevance feedback for classifying e-mail. We have theoretically given reason why our provided architecture will be unbeaten from spammers. We are now bending ourselves in simulating our architecture and its practical implementation. Although a direct comparison has not been made here, the results of SVMs for spam categorization [8] made it promising. The paper actually motivated to find a framework for spam filter using SVMs. It is a try to explore dynamic updation in learning module that seems not to be written again due to change in spam.

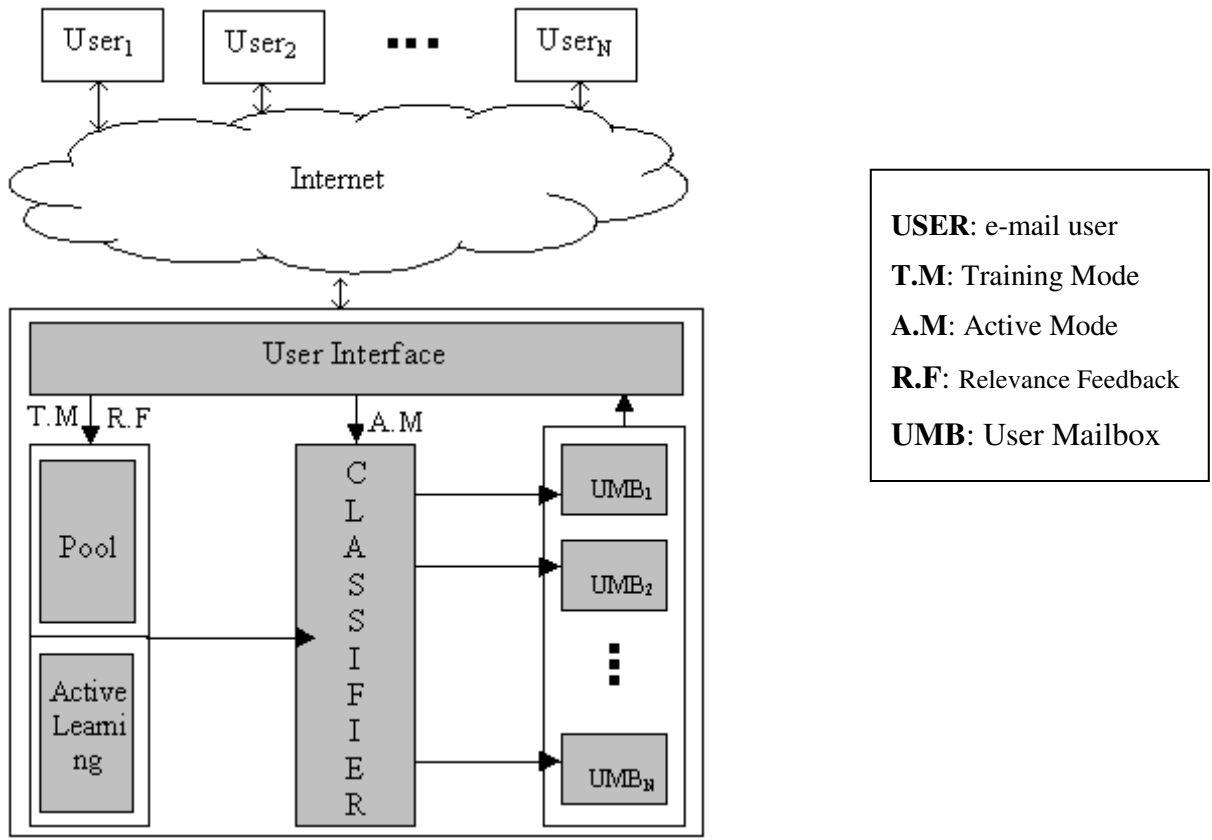

**Fig 2. Architecture of Active Learning SVMs with Relevance Feedback**